\documentclass[conference]{IEEEtran}
\usepackage[marginal]{footmisc}

\usepackage{geometry}
\usepackage{tikz}
\usepackage{circledsteps}
\usepackage{booktabs}
\usepackage{multirow}
\usepackage{multicol}
\usepackage{subfigure}
\usepackage{tikz}
\usepackage{circledsteps}
\usepackage{makecell}
\usepackage{subfigure}
\usepackage[normalem]{ulem}
\usepackage{cite}
\usepackage{amsmath,amssymb,amsfonts}
\usepackage{algorithmic}
\usepackage{graphicx}
\usepackage{textcomp}
\usepackage{xcolor}
\usepackage{soul}
\newcommand{\guangxi}[1]{\textcolor{black}{#1}}

\begin{document}
%\IEEEoverridecommandlockouts

\title{\Large\bf \vspace{-0.5cm}Late Breaking Results: Fast System Technology Co-Optimization Framework for Emerging Technology Based on Graph Neural Networks\\~\\\vspace{-0.5cm} 
	}

% \author{
%     \IEEEauthorblockN{Tianliang Ma, Xuguang Sun, Guangxi Fan, Zhihui Deng, Kainlu Low, Leilai Shao}
%     \IEEEauthorblockA{ School of mechanical and engineering, Shanghai Jiao Tong University, Shanghai, China}
%     \IEEEauthorblockA{\{matianliang,sunxuguang,guangxifan,sjtu-dzh,kainlulow,leilaishao\}@sjtu.edu.cn}
%  }


\author{
    \IEEEauthorblockN{Tianliang Ma\textdagger, Guangxi Fan\textdagger, Xuguang Sun, Zhihui Deng,  Kainlu Low, Leilai Shao*}
    \IEEEauthorblockA{Shanghai Jiao Tong University, Shanghai, China}
}

\maketitle

\makeatletter
\def\ps@IEEEtitlepagestyle{%
  \def\@oddfoot{\mycopyrightnotice}%
  \def\@evenfoot{}%
}
\makeatother
\def\mycopyrightnotice{%
  \begin{minipage}{\textwidth}
    \footnotesize
    ~ \hfill\\~\\
  \end{minipage}
  \gdef\mycopyrightnotice{}% just in case
}

\footnote{\textdagger Tianliang Ma and Guangxi Fan contribute equally. *Corresponding author: Leilai Shao (leilaishao@sjtu.edu.cn). This work was supported by the National Key Research and Development Program of China: Design Technology Co-Optimization Methodology (grant number: 2023YFB4402700) . }

{\small\bf Abstract---
\guangxi{
This paper proposes a fast system technology co-optimization (STCO) framework that optimizes power, performance, and area (PPA) for next-generation IC design, addressing the challenges and opportunities presented by novel materials and device architectures. We focus on accelerating the technology level of STCO using AI techniques, by employing graph neural network (GNN)-based approaches for both TCAD simulation and cell library characterization, which are interconnected through a unified compact model, collectively achieving over a 100X speedup over traditional methods. These advancements enable comprehensive STCO iterations with runtime speedups ranging from 1.9X to 14.1X and supports both emerging and traditional technologies.
}}
%\end{abstract}
% \begin{IEEEkeywords}
% System Technology Co-Optimization, Graph Neural Networks, TCAD Simulation, Compact Model, Cell Library Characterization
% \end{IEEEkeywords}
\vspace{-0.4cm}
\section{Introduction}
\guangxi{STCO integrates design and manufacturing early in chip development within the semiconductor industry, aiming to optimize performance, power efficiency, and yield through a unified consideration of design and manufacturing constraints. It enables concurrent exploration of materials, device structures, processes, and design choices, improving system performance and reliability. Nevertheless, the iterative nature of traditional STCO approaches prolongs development, necessitating innovative solutions to expedite the process \cite{dtco_describtion}.
}

\begin{figure}[htbp]
\centering
\includegraphics[width=0.9\linewidth]{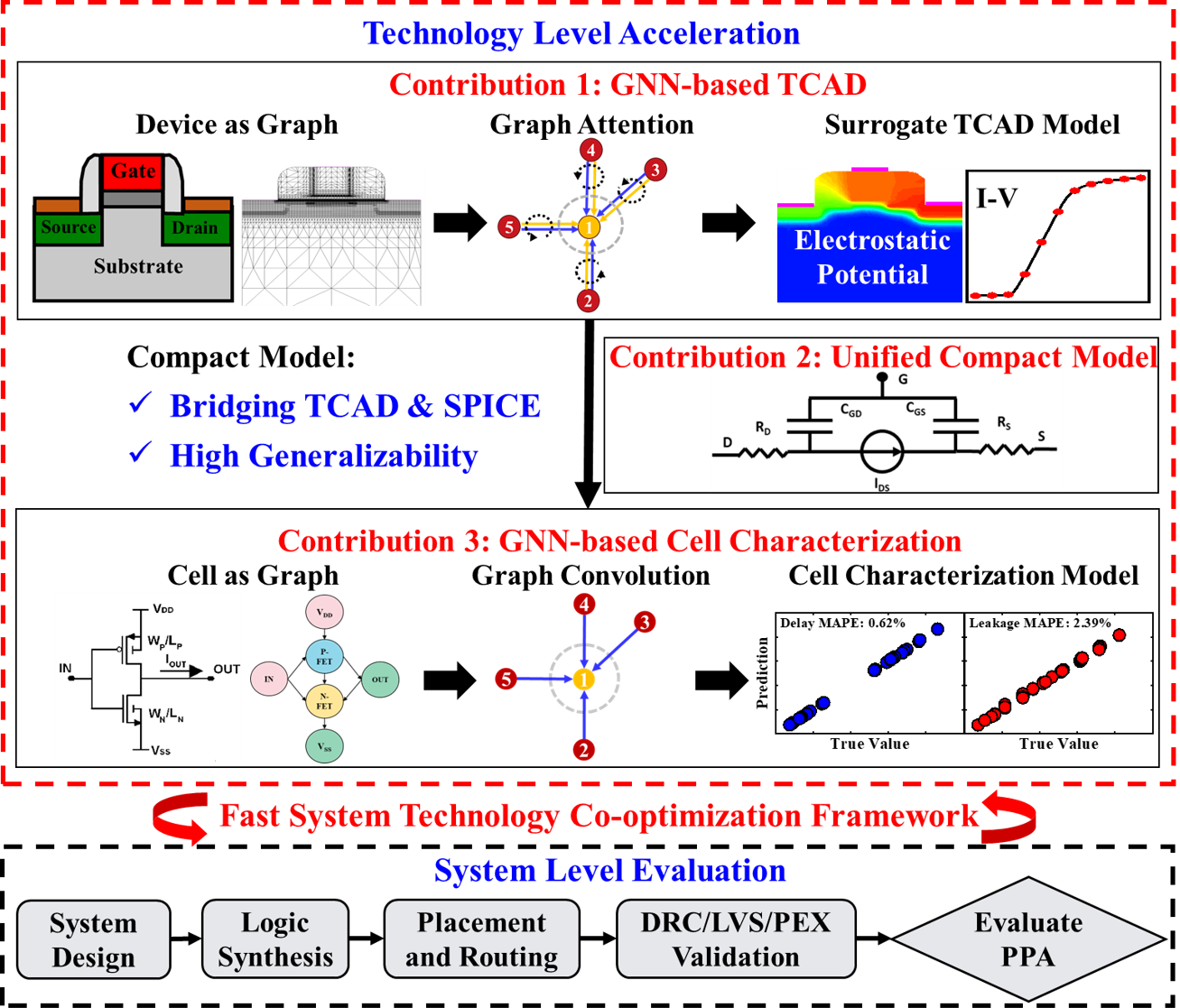}
\caption{Fast STCO framework based on GNN.}
\label{fig:intro}
\end{figure}

\guangxi{
Our work advances the acceleration of technology level tasks within the STCO domain, an area that attracts more attention in recent research. The tasks at this level often encounter significant challenges, such as time-intensive TCAD simulations for device optimization \cite{guangxi_TCAD_gnn}, the lack of standardized models in transistor modeling \cite{Compact_Model}, and the complexities in cell library characterization with emerging technologies and process variations \cite{cell library}. Fig. \ref{fig:intro} outlines our primary contributions to address these challenges, incorporating a TCAD surroage model and cell characterization model based on GNN techniques. Additionally, parameter extraction is facilitated through our unified compact model, validated across technologies such as carbon nanotube (CNT), indium gallium zinc oxide (IGZO), and low-temperature polycrystalline silicon (LTPS). Furthermore, the efficacy of our framework is validated by evaluating multiple designs using CNT technology, all conducted within the comprehensive STCO process depicted in Fig. \ref{fig:intro}.
}

\section{Fast System Technology Co-optimization Framework}
\label{Fast System Technology Co-optimization Framework}
% The greatest advantage of our fast STCO framework lies in its ability to accelerate traditional STCO process. In our practical use of this framework, we implemented a RL (reinforcement learning) agent to explore design space for ten benchmarks from ISCAS89, two MAC (multiplier accumulator) cores for AI accelerator and two open-sourced RISC-V cores. The time consumption for TCAD simulation and cell library characterization was approximately 18s for fast STCO framework while the time cost was around 2000s for commercial EDA tools, respectively. In this stage, we got over 100X acceleration. For system evaluation, we conducted logic synthesis, placement\&routing and DRC\&LVS checking using commercial tools. Finally, we compared the total runtime between our STCO framework and traditional STCO flow for one iteration, as shown in Table \ref{tab:run_time_compare_one_step}. Our fast STCO framework can achieve 1.9X to 15.7X acceleration compared to traditional optimization flow. 
\guangxi{
The framework accelerates the traditional process by employing a reinforcement learning (RL) agent to explore the design space across diverse benchmarks based on CNT technology, including six ISCAS89 benchmarks, two AI accelerator MAC (multiplier accumulator) cores, and two open-sourced RISC-V cores. Our framework significantly enhances computational efficiency, reducing TCAD simulation time to 1.38 seconds and cell library characterization time to 8.88 seconds, beyond a shared environment setup time of 8.12 seconds for both processes. In contrast, commercial EDA tools require an average of 142.07 seconds for device simulations, as determined from a calibrated study involving 576 planar CNT devices with 2D TCAD simulations\cite{cnt_tcad}, and nearly 1900 seconds for cell library characterizations. These advancements demonstrate an acceleration exceeding 100 times for both individual tasks. For system evaluation, we utilized commercial tools for logic synthesis, placement \& routing, and DRC \& LVS checks. As illustrated in Table \ref{tab:run_time_compare_one_step}, the whole flow achieves a 1.9X to 14.1X acceleration over traditional flows per iteration. The framework focuses on technology acceleration, despite relying on commercial tools for system evaluation, promises notable speed gains. With numerous AI-driven methods available to hasten system evaluation, we anticipate even greater acceleration for our STCO framework ahead. In addtion, though initially tested on CNT technology, its adaptability allows easy application to other technologies like IGZO and LTPS. %\tianliang{Considering the abundance of existing machine learning methods that can expedite the system evaluation process, the STCO framework is expected to offer even faster acceleration in the future. }
}
% In this section, we then present the details of our fast STCO framework.
\begin{table}[htbp] %one column table in two-column page
    \centering
    %\caption{Runtime Comparison between fast STCO framework and traditional optimization flow}
    \caption{Runtime Comparison between fast STCO traditional flow}
    \resizebox{\linewidth}{!}{
    \begin{tabular}{cccccc}
    \toprule
    \multirow{2}{*}{Benchmarks}& \multicolumn{3}{c}{Approximate Runtime(s)}& \multirow{2}{*}{Speedup(X)}\\
    \cline {2-4}
    \noalign{\vspace{1mm}}
    &System Evaluation &Traditional STCO Framework    &Ours\\
    \midrule
    s298&               142& 2184& 160&  13.6\\ 
    s386&               136& 2178& 154&  14.1\\
    s526&               202& 2244& 220&  10.2\\
    s820&               198& 2240& 216&  10.4\\
    s1196&              223& 2265& 241&  9.4\\
    s1488&              230& 2272& 248&  9.2\\
    16bit MAC&          536& 2578& 554&  4.7\\
    32bit MAC&          1270& 3312& 1288&  2.6\\
    Picorv32&           939& 2981& 957&  3.1\\
    Darkriscv&          2250& 4292& 2268&  1.9\\
    \bottomrule
    \end{tabular}
    %\label{tab:run_time_compare_one_step}
    }
\label{tab:run_time_compare_one_step}
\vspace{1mm}
{\footnotesize\parbox{\linewidth}{\guangxi{
Note: "Traditional STCO Framework" runtime combines system evaluation with commercial TCAD and cell library characterization times. "Ours" reflects system evaluation plus GNN-accelerated processes.}}
}
\end{table}
\subsection{GNN-based Surrogate Model for TCAD Simulation}
\begin{figure}[htb]
\centering
\includegraphics[width=0.9\linewidth]{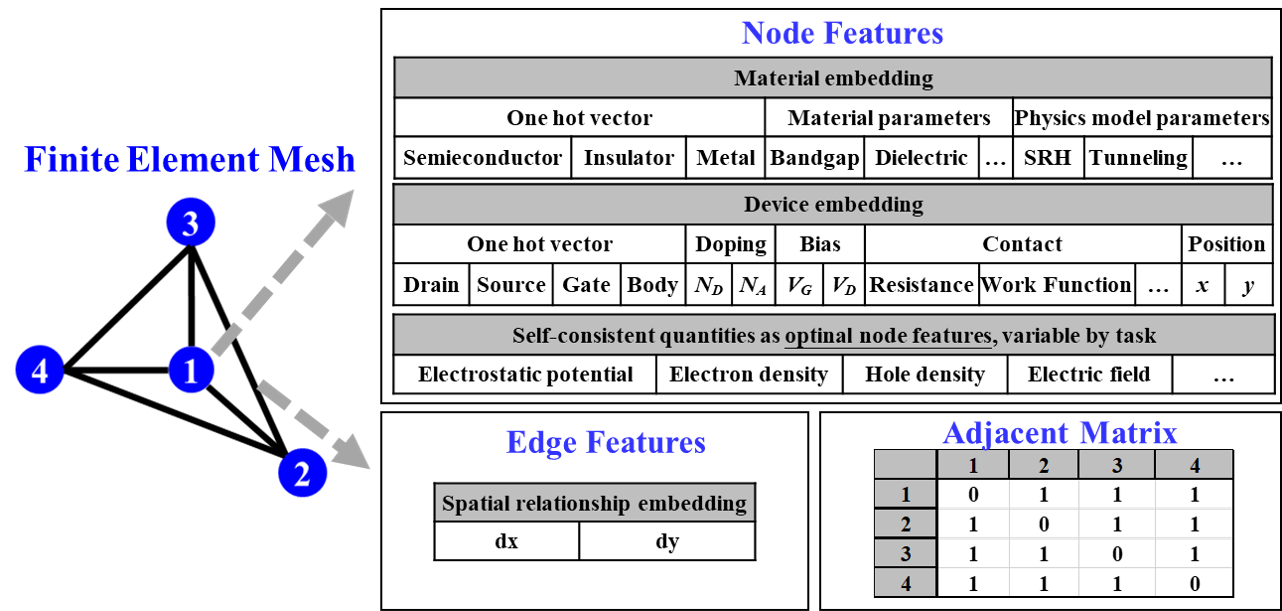}
\caption{Unified device encoding scheme based on finite element mesh details node and edge features with task-specific self-consistent quantities.}
\label{fig:GNN4TCAD_overall}
\end{figure}

\guangxi{
The graph-based unified device encoding in Fig. \ref{fig:GNN4TCAD_overall}  features material-level and device-level embeddings, alongside spatial relationship representation. Material-level embedding comprises a one-hot vector for material type delineation and a parameter vector that details material properties and parameterizes physical models, such as Shockley-Read-Hall (SRH) recombination and tunneling. Device-level embedding is designed to intricately characterize individual devices, utilizing a one-hot vector for regional traits and an attribute vector to specify the position and operational parameters, encompassing doping, bias, and more. Spatial relationship embedding, serving as edge features and inspired by finite element methods, represents the relative positions between nodes. In addition, the encoding enables inclusion of task-specific self-consistent features in device simulation like charge density,  tailored to the different tasks.
}

\guangxi{
Utilizing node and graph regression, we established a Poisson emulator and an IV predictor for device characterization respectively. The Poisson emulator integrated charge density as an additional feature, whereas the node features of the IV predictor included both charge density and potential. Architecturally, the Poisson emulator employed a deep graph attention network with edge feature (RelGAT) and comprised approximately 1 million parameters, incorporating a 12-layer GAT with 2 attention heads and one multilayer perceptron (MLP). In contrast, the IV predictor utilized a shallower RelGAT model with about 0.15 million parameters, featuring a 3-layer, single-head GAT with a 4-layer MLP for prediction. Layer normalization was applied in the training of both models, enhancing model convergence and stability across the dataset of 50,000 independent devices. Table \ref{tab:performance-GNN4TCAD} showcases the effectiveness of the RelGAT model, demonstrating its high accuracy through validation and testing. To further assess the generalizability, an additional test on 32,000 unseen data samples was conducted, indicating a highly accurate surrogate TCAD model.
}
 \begin{table}[htbt] %one column table in two-column page
    \centering
    %\caption{Performance benchmarking of MSE on dataset and an additional unseen set with 32,000 samples}
    \caption{MSE of Surrogate TCAD in the whole testing dataset}
    \resizebox{\linewidth}{!}{
    \begin{tabular}{ccccc}
    \hline
     & Validation & Testing & Unseen(32K) & $R^2(32K)$\\
    \hline
    Poisson Emulator & $6.17\times10^{-5}$ & $7.02\times10^{-5}$ & $7.15\times10^{-5}$ & 0.9999 \\
    IV Predictor & $1.67\times10^{-3}$ & $1.60\times10^{-3}$ & $1.78\times10^{-3}$ & 0.9999\\
    \hline
    %\bottomrule
    \end{tabular}
    \label{tab:performance-GNN4TCAD}
    }
\end{table}
\subsection{Unified Compact Model for Emerging Technologies}
\begin{figure}[htbp]
\centering
\includegraphics[width=0.9\linewidth]{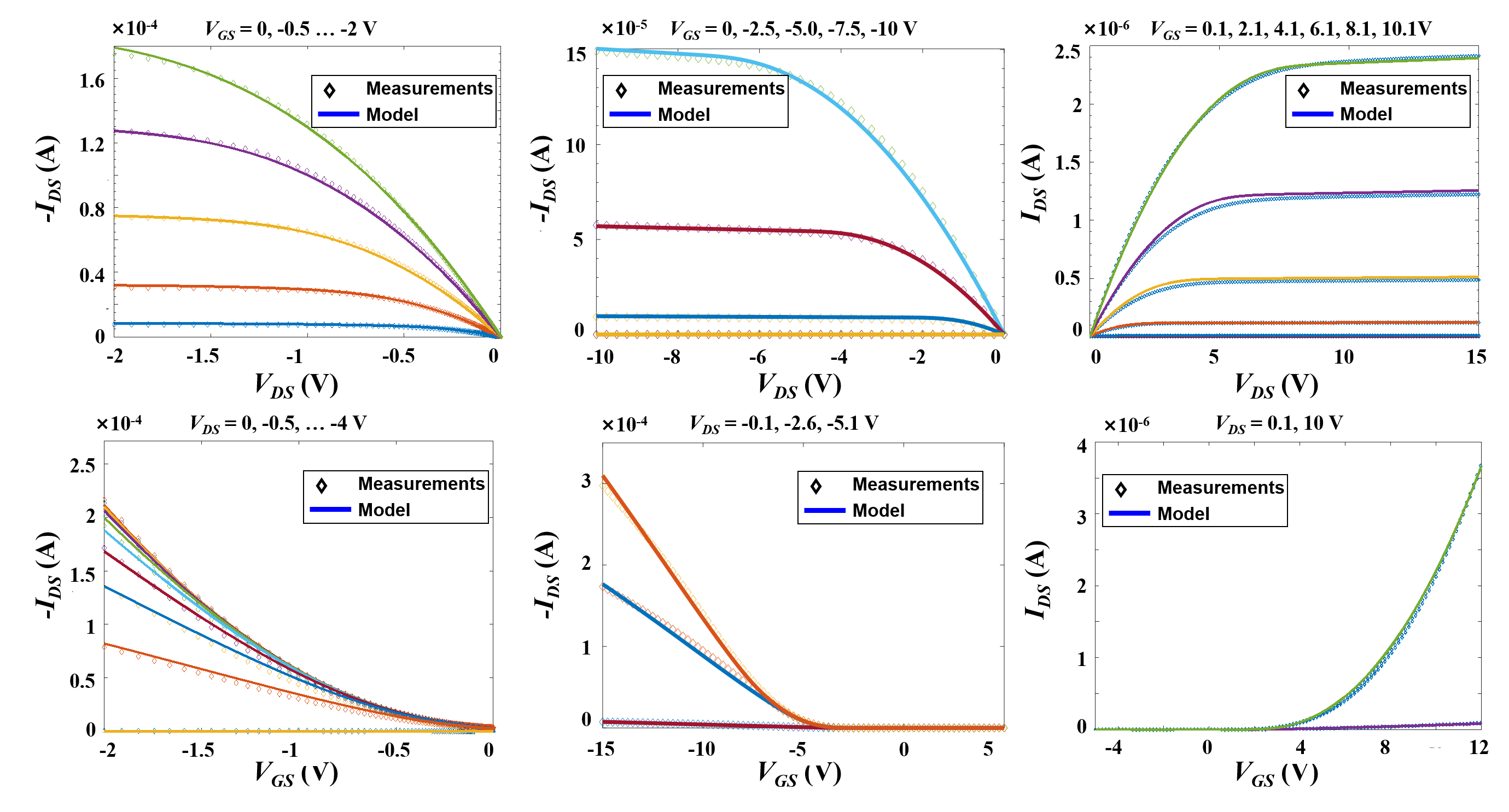}
\caption{Validations of the proposed TFT model with measured \textit{I-V} curves: (a) CNT-TFT with $L=25\mu m$ and $W=125\mu m$; (b) LTPS-TFT with $L=16\mu m$ and $W=40\mu m$; (c) IGZO-TFT with $L=20\mu m$ and $W=30\mu m$.}
\label{fig:IV_curve}
\end{figure}

%  \begin{figure}[htb]
% \centering
% \begin{minipage}[t]{0.7\linewidth}
% \includegraphics[width=1\linewidth]{figs/compact_model.png}
% \end{minipage}
% \caption{Unified compact model for emerging transistors \cite{Compact_Model}.}
% \label{fig.Model_sch}
% \end{figure} 
% The compact model is developed based on charge drift in the presence of tail-distributed traps (TDTs) and variable range hopping (VRH), which can explain the well observed mobility dependency phenomenon in carbon nanotubes (CNT), indium gallium zinc oxide (IGZO) and Low-temperature polycrystalline silicon (LTPS), as shown in Eq. (\ref{eq.mobility_dep}):
\guangxi{
The compact model for emerging transistors is formulated to account for mobility variations in CNT, IGZO, and LTPS technologies due to charge drift in the presence of tail-distributed traps (TDTs) and variable range hopping (VRH) \cite{Compact_Model}, as detailed in Eq. (\ref{eq.mobility_dep}):
}
\begin{align}
\mu= \begin{cases} \mu_0(V_{G}-V_{th})^{\gamma}, & \text{N-type TFT}\\ \mu_0(V_{th}-V_G)^{\gamma}, & \text{P-type TFT}
    \end{cases}
\label{eq.mobility_dep}
\end{align}
\guangxi{
where $V_{th}$ is the threshold voltage, $\gamma$ is the field enhancement factor for mobility and $\mu_0$ is defined as the effective mobility when $|V_{G}-V_{th}|=1$. This model integrates mobility enhancement assumptions with the charge drift to develop an intrinsic current model. Validation against measured I–V curves from CNT, LTPS, and IGZO based devices is shown in Fig. \ref{fig:IV_curve}. }
% Additionally, a depletion Pseudo-CMOS (Pseudo-D) inverter's SPICE-simulated voltage transfer curves (VTCs) closely matched actual circuit measurements across various VDD, as illustrated in Fig. \ref{fig.INVD_All}, demonstrating the model's accuracy. Despite its simplicity, the model effectively captures the primary characteristics of transistors, fulfilling our requirements for digital circuit STCO. 
% add by guangxi
\subsection{GNN-based Fast Cell Library Characterization Model}
Our methodology employs a two-stage process to advance cell characterization for semiconductor design. Initially, we developed a comprehensive cell library comprising 35 types of combinational and sequential cells and utilized transistor-level SPICE simulation to generate extensive cell datasets for training and testing, encompassing nine metrics including delay, output slew, leakage power, capacitance (maximum capacitance of each input pin), flip power (dynamic power generated when both input and output pins are flipped), non-flip power (dynamic power generated when only the input pins are flipped, while the state of the output pin remains unchanged), minimum setup/hold time, and pulse width (for sequential cells only) under various corners and technology settings. In the second stage, we adopted a \guangxi{3-layer} graph convolutional network (GCN) to establish our framework. To enhance the accuracy of predictions, \guangxi{an additional 2-layer MLP} was added after the GCN layers for each metric. 

In our study of emerging technology, we utilized the unified compact model and specifically focused on analyzing the variation of supply voltage ($V_{DD}$), threshold voltage ($V_{th}$), and gate unit capacitance ($C_{ox}$). These three critical parameters significantly influences the performance of emerging devices. The definitions of node features are presented in Table \ref{tab:node_features}. Here, "Current\_state" and "Next\_state" are two features related to the input pin state, with "1" denoting a high level and "0" representing a low level, respectively. "Input\_slew" denotes the transition time of input signal, "Output\_load" denotes capacitive load on cell output pins. For cell metrics that do not have relationship with some bits in the vector, these bits will be set to ”0”.
\begin{table}[htbp] %one column table in two-column page
    \centering
    \caption{Node feature vector definition for silicon technology }
    \resizebox{\linewidth}{!}{
    \begin{tabular}{ccccccc}
    %\toprule
    \hline
    &  IN& OUT& N-FET& P-FET& $V_{DD}$& $V_{SS}$\\
    \hline
    %\midrule
     Bit0&       0& 0& 0& 0& 1& 1\\     
     Bit1&       0& 1& 1& 1& 0& 0\\
     Bit2&       1& 0& 1& 1& 0& 1\\
     Bit3&       0& 0& -1& 1& 0& 0\\
     Bit4&       0& 0& 0& 0& $V_{DD}$& 0\\
     Bit5&       0& 0& Width& Width& 0& 0\\
     Bit6&       0& 0& Gate Unit Capacitance& Gate Unit Capacitance& 0& 0\\
     Bit7&       0& 0& $V_{th}$& $V_{th}$& 0& 0\\
     Bit8&       Input\_slew& 0& 0& 0& 0& 0\\
     Bit9&       0& Output\_load& 0& 0& 0& 0\\
     Bit10&      Current\_state& 0& 0& 0& 0& 0\\
     Bit11&      Next\_state& 0& 0& 0& 0& 0\\
    \hline
    %\bottomrule
    \end{tabular}
    \label{tab:node_features}
    }
\end{table}

As detailed in Table \ref{tab:MAPEs_test}, valid data from 125 corners were for training while from 512 corners for testing, the model demonstrated high accuracy in experiments. 
For flip power and non-flip power, the average error of prediction is relatively larger compared with other cell metrics, the reason is due to the fact that the dynamic power consumption varies by several orders of magnitude among different standard cells. The model exhibits a relatively large percentage error in predicting extremely low dynamic power consumption which needs to be improved in our future work.
% In the second stage, the model undergoes validation and is integrated into a fast STCO framework, aiming to optimize the design and performance of semiconductor devices efficiently. 
%\tianliang{
%We've built a thorough cell library with 35 combinational and sequential cell types, assessing nine metrics based on input transition time and effective capacitive load using a non-linear delay model (NLDM). Employing the Graph Convolutional Network (GCN), a classic GNN method, we established our framework. With 125 corners in the training dataset and 512 in testing, our model demonstrates high accuracy in experiments, detailed in Table \ref{tab:MAPEs_test}.}
\begin{table}[htbp] %one column table in two-column page
    \centering
    \caption{MAPEs of cell library prediction in the whole testing dataset}
    \resizebox{\linewidth}{!}{
    \begin{tabular}{cccc}
    %\toprule
    \hline
    & LTPS& CNT& Number of Data Points\\
    \hline
    %\midrule
    Delay&  0.47\%& 0.62\%& 696320\\ 
    Output Slew&  0.79\%& 0.83\%& 696320\\
    Capacitance& 0.18\%& 0.21\%& 70656\\
    Flip Power& 5.74\%& 4.96\%& 696320\\
    Non-flip Power& 3.36\%& 5.60\%& 393216\\
    Leakage Power& 2.78\%& 2.39\%& 165888\\
    Minimum Pulse Width& 1.20\%& 1.67\%& 8192\\
    Minimum Setup& 0.50\%& 0.27\%& 16384\\
    Minimum Hold& 0.45\%& 0.38\%& 16384\\
    \hline
    %\bottomrule
    \end{tabular}
    \label{tab:MAPEs_test}
    }
\end{table}
% \section*{\sc Acknowledgments}
% %\begin{acks}
% This material is based upon work supported, in part, by grant from Shanghai Sailing Program (No. 22YF1420100). 
% %\end{acks}

% The next two lines define the bibliography style to be used, and
% the bibliography file.
%\bibliographystyle{ACM-Reference-Format}
%\bibliography{sample-base}

% \appendix

\end{document}